\documentclass[conference,a4paper]{IEEEtran} %for SPL
\IEEEoverridecommandlockouts
\usepackage[binary-units,per-mode=symbol,per-symbol=p]{siunitx}
\usepackage{cite}
\usepackage{adjustbox}
\usepackage[ruled,vlined]{algorithm2e}
\usepackage{amsmath,amssymb,amsfonts}
\usepackage{algorithmic}
\usepackage{graphicx}
\usepackage{textcomp}
\usepackage[colorlinks=false, pdfborder={0 0 0}, breaklinks]{hyperref}
\usepackage{xcolor}
\usepackage{balance}
\usepackage{enumitem}
\setlist{nolistsep,leftmargin=*}
\usepackage{setspace}
\usepackage{wrapfig}
\usepackage{listings}
\usepackage{color}
\usepackage{caption}
\usepackage{subcaption}
\usepackage{todonotes}
\usepackage[longtable]{multirow}

\def\BibTeX{{\rm B\kern-.05em{\sc i\kern-.025em b}\kern-.08em
    T\kern-.1667em\lower.7ex\hbox{E}\kern-.125emX}}

\usepackage{amsmath}

\usepackage{ctable}

\newcommand{\etal}{\emph{et al.}\xspace}
\newcommand{\ie}{\emph{i.e.}, }
\newcommand{\eg}{\emph{e.g.}, }

\newcommand{\cf}{\emph{cf.}\xspace}
\newcommand{\HAS}{\emph{HTTP Adaptive Streaming}\xspace}

\newcommand{\VOD}{\emph{Video on Demand }}

\newcommand{\scheme}{\texttt{DECODRA}\xspace}

\newcommand{\EY}{$E_{\text{Y}}$}
\newcommand{\h}{$h$}
\newcommand{\LY}{$L_{\text{Y}}$}

\begin{document}
%\bstctlcite{IEEEexample:BSTcontrol}
\title{Energy-Quality-aware Variable Framerate Pareto-Front for Adaptive Video Streaming}

%\author{Anonymous VCIP Submission}
\author{\IEEEauthorblockN{Prajit T Rajendran\IEEEauthorrefmark{1},  Samira Afzal\IEEEauthorrefmark{2}, Vignesh V Menon\IEEEauthorrefmark{3},  Christian Timmerer\IEEEauthorrefmark{2}
\thanks{The financial support of the Austrian Federal Ministry for Digital and Economic Affairs, the National Foundation for Research, Technology and Development, and the Christian Doppler Research Association is gratefully acknowledged. Christian Doppler Laboratory ATHENA: \url{https://athena.itec.aau.at/}.}
}
\IEEEauthorblockA{
\IEEEauthorrefmark{1}\small Universite Paris-Saclay, CEA, List, F-91120, Palaiseau, France\\
\IEEEauthorrefmark{2}\small Christian Doppler Laboratory ATHENA, Institute of Information Technology (ITEC), Alpen-Adria-Universität, Klagenfurt, Austria\\
\IEEEauthorrefmark{3}\small Video Communication and Applications Department, Fraunhofer HHI, Berlin, Germany\\
\vspace{-2.6em}
}
}

\maketitle

\begin{abstract}
Optimizing framerate for a given bitrate-spatial resolution pair in adaptive video streaming is essential to maintain perceptual quality while considering decoding complexity. Low framerates at low bitrates reduce compression artifacts and decrease decoding energy. We propose a novel method, \underline{Decod}ing-complexity aware F\underline{ra}merate Prediction (\scheme), which employs a Variable Framerate Pareto-front approach to predict an optimized framerate that minimizes decoding energy under quality degradation constraints. \scheme dynamically adjusts the framerate based on current bitrate and spatial resolution, balancing trade-offs between framerate, perceptual quality, and decoding complexity. Extensive experimentation with the \mbox{Inter-4K} dataset demonstrates \scheme's effectiveness, yielding an average decoding energy reduction of up to \SI{13.45}{\percent}, with minimal VMAF reduction of 0.33 points at a low-quality degradation threshold, compared to the default 60 fps encoding. Even at an aggressive threshold, \scheme achieves significant energy savings of \SI{13.45}{\percent} while only reducing VMAF by \SI{2.11}{} points. In this way, \scheme extends mobile device battery life and reduces the energy footprint of streaming services by providing a more energy-efficient video streaming pipeline.
\end{abstract}


\maketitle

\setlength{\textfloatsep}{1pt}

\section{Introduction}
In today's digital age, \HAS~(HAS)\cite{DASH_Survey} has revolutionized multimedia consumption by providing seamless access to high-quality video content across various devices, including smartphones, tablets, and smart TVs\cite{uhdtv_ref, Sodagar2011}. The increasing proliferation of high-speed internet and the widespread adoption of connected devices have heightened user expectations for uninterrupted, high-quality video streaming~\cite{uhdtv_ref}. HAS addresses these demands by dynamically adjusting video quality based on network bandwidth, device capabilities, and user preferences~\cite{DASH_ref}, optimizing encoding parameters such as bitrate, spatial resolution, and framerate to ensure the best viewing experience. In this paper, we define spatial resolution as the pixel dimensions (width × height) of the video and temporal resolution as the framerate in frames per second (fps). Each video segment is encoded in multiple codec representations for every supported codec~\cite{mcbe_ref}, enabling adaptive streaming clients to switch between quality levels as network conditions fluctuate~\cite{emes_ref}.

\begin{figure}[t]
\centering
\includegraphics[width=0.491\columnwidth]{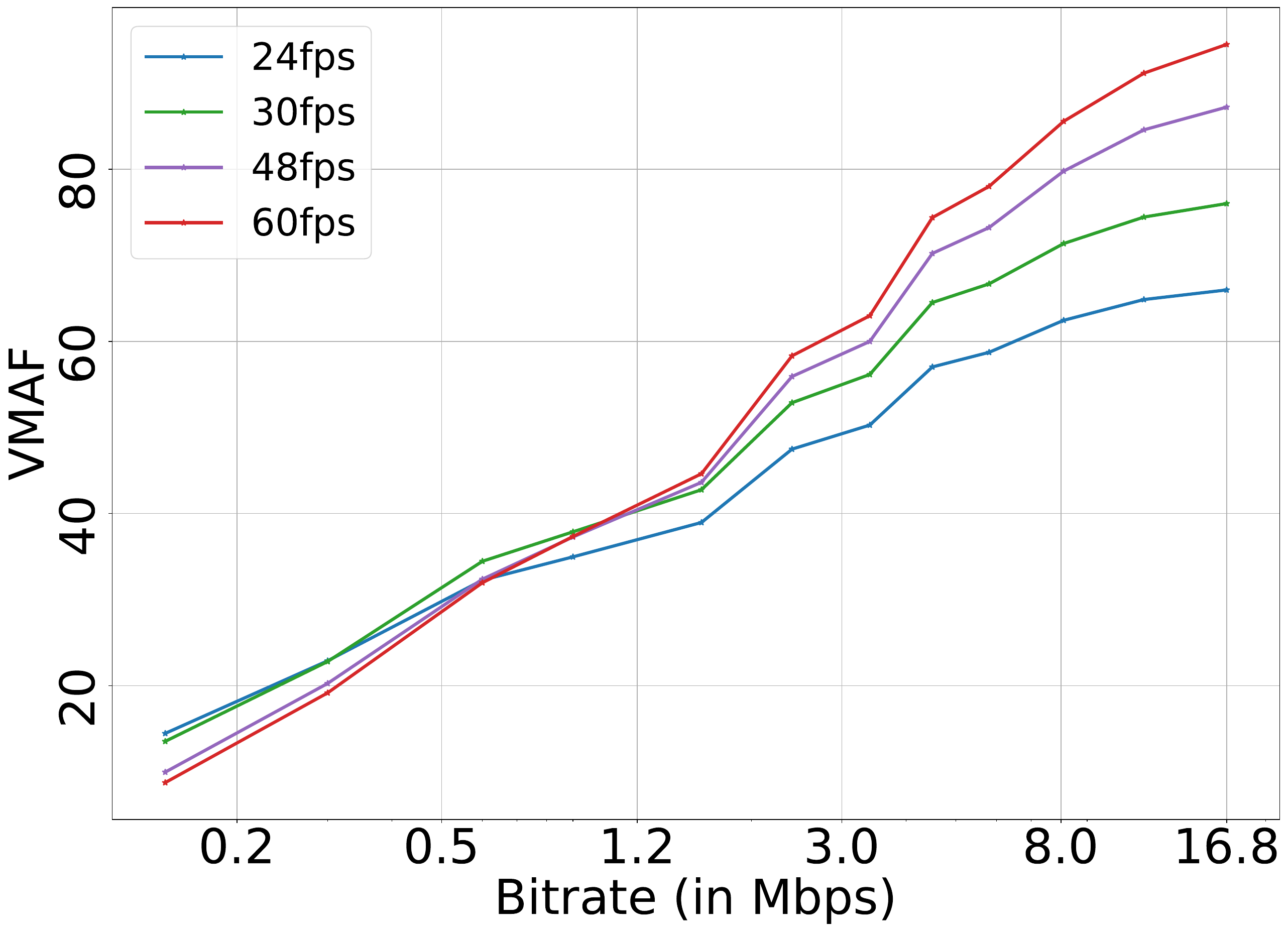}
\includegraphics[width=0.491\columnwidth]{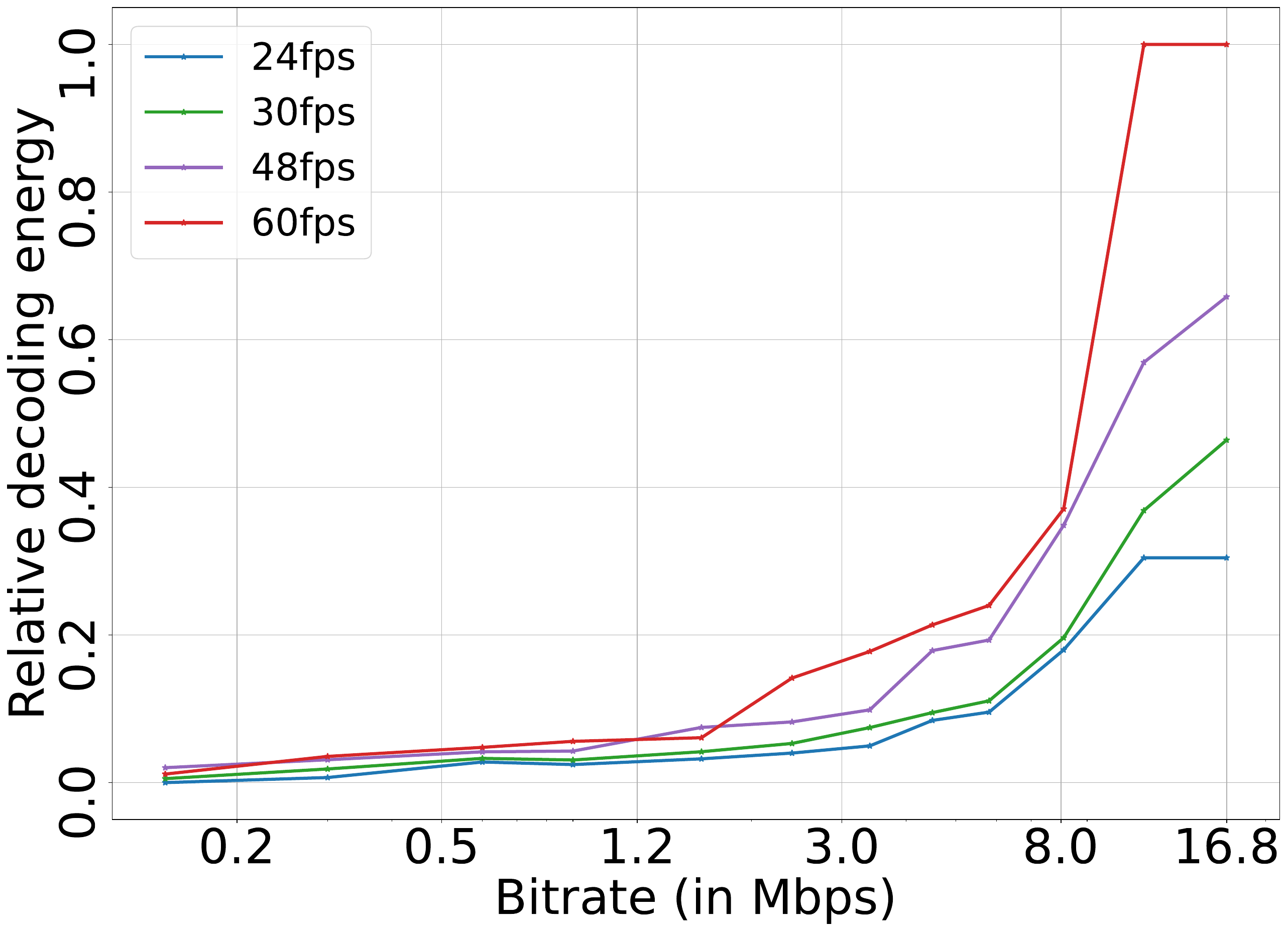}
%\vspace{-0.74em}
\caption{Rate-VMAF and rate-decoding energy for multiple framerates for Sequence \emph{0286} of Inter-4K dataset~\cite{inter4k_ref} encoded using x265 encoder at \emph{veryslow} preset for Apple HLS ladder~\cite{HLS_ladder_ref}.}
\label{fig:intro}
\end{figure}

Framerate, the number of frames displayed per second (fps), is critical in ensuring motion smoothness and visual quality~\cite{nasiri_2015, mackin_hfr_motion}. Higher framerates (HFR) are increasingly popular in broadcast~\cite{hfr_tv}, online streaming (e.g., YouTube supports up to 60\,fps), and gaming communities~\cite{hfr_gaming_ref}. The UHDTV standard specifies framerates up to 120\,fps~\cite{uhdtv_ref}. However, increasing the framerate also increases computational complexity during video decoding (\cf Fig.~\ref{fig:intro}), posing challenges for resource-constrained devices such as smartphones and tablets~\cite{Herglotz_2019}. Decoding complexity, defined as the computational resources required for real-time video frame decoding, is a crucial consideration in adaptive streaming, as it impacts device battery life, CPU usage, and overall user experience. In \VOD (VOD) platforms, videos are encoded once but decoded multiple times on client devices, making decoding efficiency increasingly important as viewership grows~\cite{katsenou2022energy, katsenou2024ratequality}. Reducing decoding energy consumption is, therefore, vital for improving the overall energy efficiency of the streaming pipeline~\cite{dec_comp_vvc_ref}.

While bitrate and spatial resolution optimization have been extensively studied~\cite{gnostic, netflix_paper, jtps_ref, katsenou2024ratequality}, framerate optimization has received less attention despite its significant impact on perceptual quality, computational complexity, and overall streaming performance. The Variable Framerate (VFR) Pareto-front (PF) approach presents substantial advantages over traditional dynamic spatial resolution PF methods, particularly in terms of Quality of Experience (QoE) and decoding complexity. Dynamic spatial resolution methods~\cite{katsenou2024ratequality, decode_energy_vvc_ref, decode_energy_vvc_ref1} adjust the spatial resolution to balance quality and resource usage, but they can result in noticeable shifts in visual detail and clarity. These changes can disrupt the viewer's experience, especially when switching between high and low spatial resolutions. In contrast, the VFR approach fine-tunes the framerate, allowing for more subtle adjustments that preserve the consistency of visual quality. This leads to a smoother viewing experience with fewer perceptual disturbances. It is also noteworthy that traditional framerate optimization approaches often neglect decoding complexity, leading to inefficiencies on low-power devices~\cite{vfr_csvt2_ref, vfr_pcs1_ref, vfrc_ref, cvfr_ref}. Our proposed decoding complexity-aware approach (\scheme) addresses this gap by dynamically adjusting the framerate to minimize decoding energy consumption while maintaining high-quality video playback. \scheme optimally balances visual quality and computational resources, ensuring smooth and efficient streaming across various devices.

%The remainder of the paper is organized as follows. Section~\ref{sec:fr_opt} discusses optimizing framerate in the context of adaptive video streaming, while Section~\ref{sec:decodra} explains our proposed approach and implementation details and discusses the trade-offs in optimizing perceptual quality, bitrate, resolution, and framerate. Section~\ref{sec:results} shows the experimental results of \scheme compared to the state-of-the-art methods, while Section~\ref{sec:concl} concludes the paper.

\section{Framerate Optimization in Adaptive Video Streaming}
\label{sec:fr_opt}
Framerate optimization is crucial in adaptive video streaming, impacting perceptual quality and streaming efficiency. Traditionally, video streams use fixed framerates, often based on source content or standard formats (\eg 24\,fps for film, 30\,fps or 60\,fps for TV)\cite{ntsc_ref1}. However, adaptive streaming can dynamically adjust framerates based on network conditions, device capabilities, and user preferences. Fixed framerate approaches select framerates according to the target device's display capabilities and bandwidth, using lower framerates for mobile devices and higher framerates for large-screen, high-spatial-resolution devices\cite{nokia_ref}.

Variable framerate (VFR) encoding dynamically adjusts framerates during playback based on real-time conditions. Techniques for VFR include thresholding motion-related features~\cite{vfr_csvt2_ref} or using machine learning algorithms~\cite{vfr_tb1_ref, vfr_pcs1_ref}. Huang~\etal \cite{vfr_tb1_ref} proposed a support vector regression method for framerate selection to meet user satisfaction ratios. Katsenou~\etal \cite{vfr_pcs1_ref} used decision trees to predict critical framerates at the sequence level using optical flow and gray-level co-occurrence matrices. Herrou~\etal \cite{vfrc_ref} used random forest classifiers to determine the minimum framerate that preserves perceived video quality. VFR encoding maximizes perceptual quality while minimizing bandwidth usage and decoding complexity but requires sophisticated algorithms and real-time adaptation for smooth playback transitions.

Existing methods overlook decoding complexity considerations, leading to suboptimal performance and inefficiencies, especially on resource-constrained devices. High decoding complexity can lead to playback stutters, frame drops, and increased battery drain on mobile devices, negatively impacting the user experience.

\section{Decoding Complexity-Aware Framerate Optimization (\scheme)}
\label{sec:decodra}

\subsection{Optimization problem formulation}
\label{sec:opt_problem}
The objective of the optimization problem is to minimize the absolute difference between the perceptual quality at a given framerate and the target quality \((v_{\text{max}} - v_{\text{J}})\), where $v_{\text{J}}$ represents the quality degradation threshold, illustrating the acceptable quality degradation level, $v_{\text{max}}$ denotes the maximum quality for the current bitrate-spatial resolution pair. However, the perceptual quality at the selected framerate must be greater than or equal to \((v_{\text{max}} - v_{\text{J}})\). Mathematically, the objective function and constraint can be expressed as:

\begin{equation}
\begin{aligned}
& \text{Objective:} && \min_{f \in \mathcal{F}} \left| v_{(b, r, f)} - (v_{\text{max}} - v_{\text{J}}) \right| \\
& \text{Constraint:} && v_{(b, r, f)} \geq v_{\text{max}} - v_{\text{J}}
\end{aligned}
\end{equation}
where $f$ represents the framerate, \ie the number of frames per second (fps), while $r$ denotes the spatial resolution of the video in terms of width × height (pixels). We can generalize the formulation to handle different levels of quality degradation thresholds by considering the following optimization problem for any \(v_{\text{J}} \geq 0\):

\begin{align}
\resizebox{0.99\hsize}{!}{
$f_{v_{\text{J}}} = \begin{cases}
\arg\max_{f \in \mathcal{F}} v_{(b, r, f)}, & \text{if } v_{\text{J}} = 0 \\
\arg\min_{f \in \mathcal{F}} \left\{ \left| v_{(b, r, f)} - (v_{\text{max}} - v_{\text{J}}) \right| : v_{(b, r, f)} \geq v_{\text{max}} - v_{\text{J}} \right\}, & \text{if } v_{\text{J}} > 0
\end{cases}$
}
\end{align}

\begin{figure}[t]
\centering
    \includegraphics[trim=0cm 0cm 26.8cm 0cm, clip, width=0.849\columnwidth]{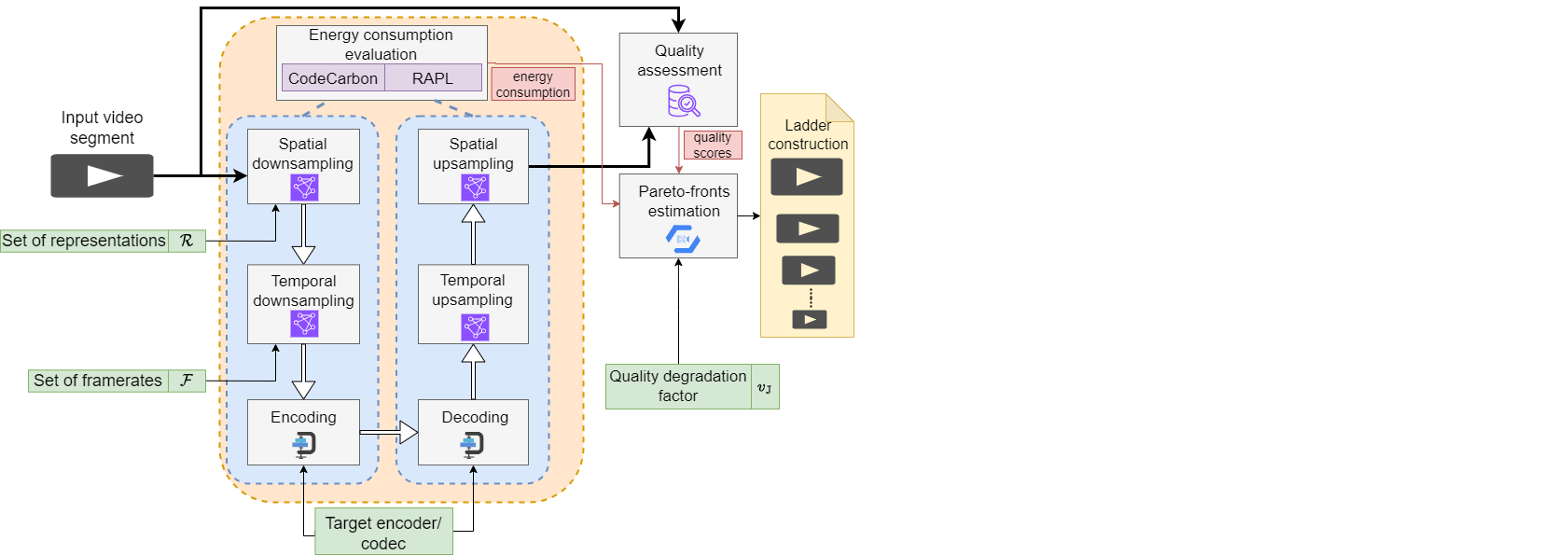}
%\vspace{-0.54em}    
\caption{\scheme implementation.}
%\vspace{-0.1em}
\label{fig:prop_method_arch}    
\end{figure}

\subsection{Implementation}
The proposed implementation involves a systematic spatial and temporal downsampling, encoding, decoding, and quality assessment process, as shown in Fig.~\ref{fig:prop_method_arch}.

\subsubsection{Spatial downsampling} involves resizing each video frame to a lower spatial resolution using interpolation techniques. In this paper, we employ the bicubic filter of FFmpeg for this purpose~\cite{ffmpeg_ref}.

\subsubsection{Temporal downsampling} is performed by selectively removing frames from the video to reduce the framerate (fps). We use frame dropping as the method for temporal downsampling, where specific frames are skipped to decrease the framerate~\cite{drop_ref1}.

\subsubsection{Encoding}
The downsampled video segments are then encoded using the target encoder with the desired preset to ensure high compression efficiency. We encode each bitrate-spatial resolution pair ($b,r$) at multiple framerates in $\mathcal{F}$.

\subsubsection{Temporal upsampling} interpolates additional frames to return the framerate to the desired level. We employ motion-compensated interpolation techniques to generate intermediate frames, ensuring smooth playback~\cite{interpolation_ref2, interpolation_ref3}.

\subsubsection{Spatial upsampling} involves resizing each video frame back to the original spatial resolution using the same bicubic filter used for downsampling. This step restores the video to its initial spatial resolution. 

\subsubsection{Decoding}
The encoded video segments are decoded using the target decoder, which simulates playback on a typical client device. The decoded output is then analyzed for quality and energy consumption. 

\subsubsection{Quality and energy measurement}
We evaluate the quality of the processed videos using objective metrics such as Peak Signal-to-Noise Ratio (PSNR)~\cite{psnr_ref1}, Structural Similarity Index (SSIM)~\cite{ssim_ref2}, and Video Multi-Method Assessment Fusion (VMAF)~\cite{VMAF}. 
We measure the decoding complexity using the Running Average Power Limit (RAPL) interface and the CodeCarbon tool~\cite{codecarbon_ref}. These tools provide accurate measurements of the energy consumed during the decoding process, allowing us to assess the energy efficiency of different framerate configurations. Energy measurements are conducted in real-time during decoding on the server. Each decoding instance is repeated three times to ensure consistency, and the energy consumption is averaged across all runs.

\subsubsection{Pareto-front estimation}
Algorithm~\ref{alg:fr_select} illustrates the PF estimation algorithm for framerate optimization, which is designed to select the optimal framerate that balances video quality and decoding energy consumption. The algorithm takes as inputs a target bitrate-spatial resolution pair $(b,r)$, a dictionary of quality values for each framerate, and a quality degradation threshold $v_{\text{J}}$. It begins by calculating the maximum quality value $v_{\text{max}}$  across all possible framerates for the given bitrate-spatial resolution pair. An optimal framerate variable $\hat{f}_t$ is then initialized to zero. 
The algorithm iterates over each framerate in the set of possible framerates ($\mathcal{F}$). For each framerate, we check if the difference between $v_{\text{max}}$ and the quality value at that framerate is within the acceptable quality degradation threshold 
$v_{\text{J}}$. If the quality constraint is satisfied and the current framerate is either the first valid framerate found or smaller than the previously selected framerate, the optimal framerate variable $\hat{f}_t$ is updated to the current framerate. This ensures that the chosen framerate minimizes the absolute difference between the quality and target quality values ($v_{\text{max}} - v_{\text{J}}$) while maintaining the quality above the threshold. The algorithm dynamically adapts the framerate to provide the best possible perceptual quality with optimized energy consumption. This paper considers VMAF~\cite{VMAF} as the perceptual quality metric for PF optimization.

\begin{algorithm}[t]
\caption{\small{PF estimation for framerate optimization.}}
\label{alg:fr_select}
\footnotesize
\begin{algorithmic}[1]
\REQUIRE $(b, r)$: Target bitrate-spatial resolution pair
\REQUIRE $v_{(b, r, f)}$: Dictionary of quality values for each framerate
\REQUIRE $v_{\text{J}}$: Quality degradation threshold
\ENSURE $\hat{f}_t$: Selected optimal framerate

\STATE $v_{\text{max}} = \max_{f \in \mathcal{F}} v_{(b, r, f)}$ \label{alg:line1}
\STATE $\hat{f}_t \gets 0$ \label{alg:line2}

\FOR{each $f \in \mathcal{F}$} \label{alg:line3}
    \IF{$v_{\text{max}} - v_{(b, r, f)} \leq v_{\text{J}}$} \label{alg:line4}
        \IF{$\hat{f}_t == 0$ \OR $f < \hat{f}_t$} \label{alg:line5}
            \STATE $\hat{f}_t = f$ \label{alg:line6}
        \ENDIF
    \ENDIF
\ENDFOR

\RETURN $\hat{f}_t$

\end{algorithmic}
\end{algorithm}

\section{Evaluation setup}
\subsection{State-of-the-art}
\begin{enumerate}
    \item \textit{Default}: adopts a fixed framerate for the CBR encoding of the HLS bitrate ladder~\cite{HLS_ladder_ref} at 60\,fps.
    \item \textit{HQ}~\cite{netflix_paper} determines the optimized framerate yielding the highest possible perceptual quality (in terms of VMAF) for each representation by encoding all framerates.
\end{enumerate}

\subsection{Dataset and metrics}
We utilize the Inter-4K dataset~\cite{inter4k_ref} with video segments representing various spatiotemporal complexities determined using the Video Complexity Analyzer tool~\cite{vca_ref} to ensure the robustness and generalizability of our results. The experimental parameters used in this paper are reported in Table~\ref{tab:exp_par}. We consider a set of video representations with varying spatial resolutions and bitrates according to Apple HLS authoring specification~\cite{HLS_ladder_ref}. We evaluate four framerates representing the National Television Standards Committee (NTSC)~\cite{ntsc_ref1} framerate options in adaptive streaming systems. Based on the typical just noticeable difference (JND)~\cite{jnd_ref,zhu2022framework} thresholds discussed in the literature, we consider quality degradation thresholds of one, two~\cite{kah_ref}, four, and six~\cite{jnd_streaming} VMAF points. We run experiments on a dual server with Intel Xeon Gold 5218R processors (80 cores, operating at 2.10 GHz).

\begin{table}[t]
\caption{Experimental parameters of \scheme used in this paper.}
\centering
\resizebox{0.815\linewidth}{!}{
\begin{tabular}{l|c|c|c|c|c|c|c}
\specialrule{.12em}{.05em}{.05em}
\specialrule{.12em}{.05em}{.05em}
\multicolumn{2}{c|}{\emph{Parameter}} & \multicolumn{6}{c}{\emph{Values}}\\
%Representation ID & 01 & 02 & 03 & 04 & 05 & 06 & 07 & 08 & 09 \\
\specialrule{.12em}{.05em}{.05em}
\specialrule{.12em}{.05em}{.05em}
\multirow{4}{*}{$\mathcal{R}$} & \emph{r [pixels]} & 360 & 432 & 540 & 540 & 540 & 720 \\
& \emph{b [Mbps]} &  0.145 & 0.300 & 0.600 & 0.900 & 1.600 & 2.400 \\
\cmidrule(lr){2-8}
& \emph{r [pixels]} & 720 & 1080 & 1080 & 1440 & 2160 & 2160 \\
& \emph{b [Mbps]} &  3.400 & 4.500 & 5.800 & 8.100 & 11.600 & 16.800\\
\hline
\multicolumn{2}{c|}{$\mathcal{F}$} & \multicolumn{6}{c}{\{24, 30, 48, 60\} } \\
\hline
\multicolumn{2}{c|}{$v_{\text{J}}$} & \multicolumn{2}{c|}{2} & \multicolumn{2}{c|}{4} & \multicolumn{2}{c}{6} \\
\hline
\multicolumn{2}{c|}{\emph{Encoder}} & \multicolumn{6}{c}{\emph{x265 v3.6 [veryslow]}}\\
\hline
\multicolumn{2}{c|}{\emph{Decoder}} & \multicolumn{6}{c}{\emph{HM}}\\
\specialrule{.12em}{.05em}{.05em}
\specialrule{.12em}{.05em}{.05em}
\end{tabular}
}
\label{tab:exp_par}
\end{table}

We evaluate Bjøntegaard Delta PSNR (\mbox{BD-PSNR}) and  Bjøntegaard Delta VMAF (\mbox{BD-VMAF})~\cite{DCC_BJDelta} measures, which refer to the average increase in quality of the representations compared to the reference bitrate ladder encoding scheme at the same bitrate. A positive value suggests a boost in the coding efficiency of the considered encoding method compared to the reference encoding method.  Moreover, we also evaluate Bjøntegaard Delta Decoding Energy (\mbox{BDDE})~\cite{Herglotz_2019}, which measures energy savings as a percentage for the same video quality. A negative \mbox{BDDE} indicates
that the same quality is achieved with reduced energy consumption.

\begin{figure}[t]
\centering
\begin{subfigure}{0.491\columnwidth}
    \centering
    \includegraphics[clip,width=\textwidth]{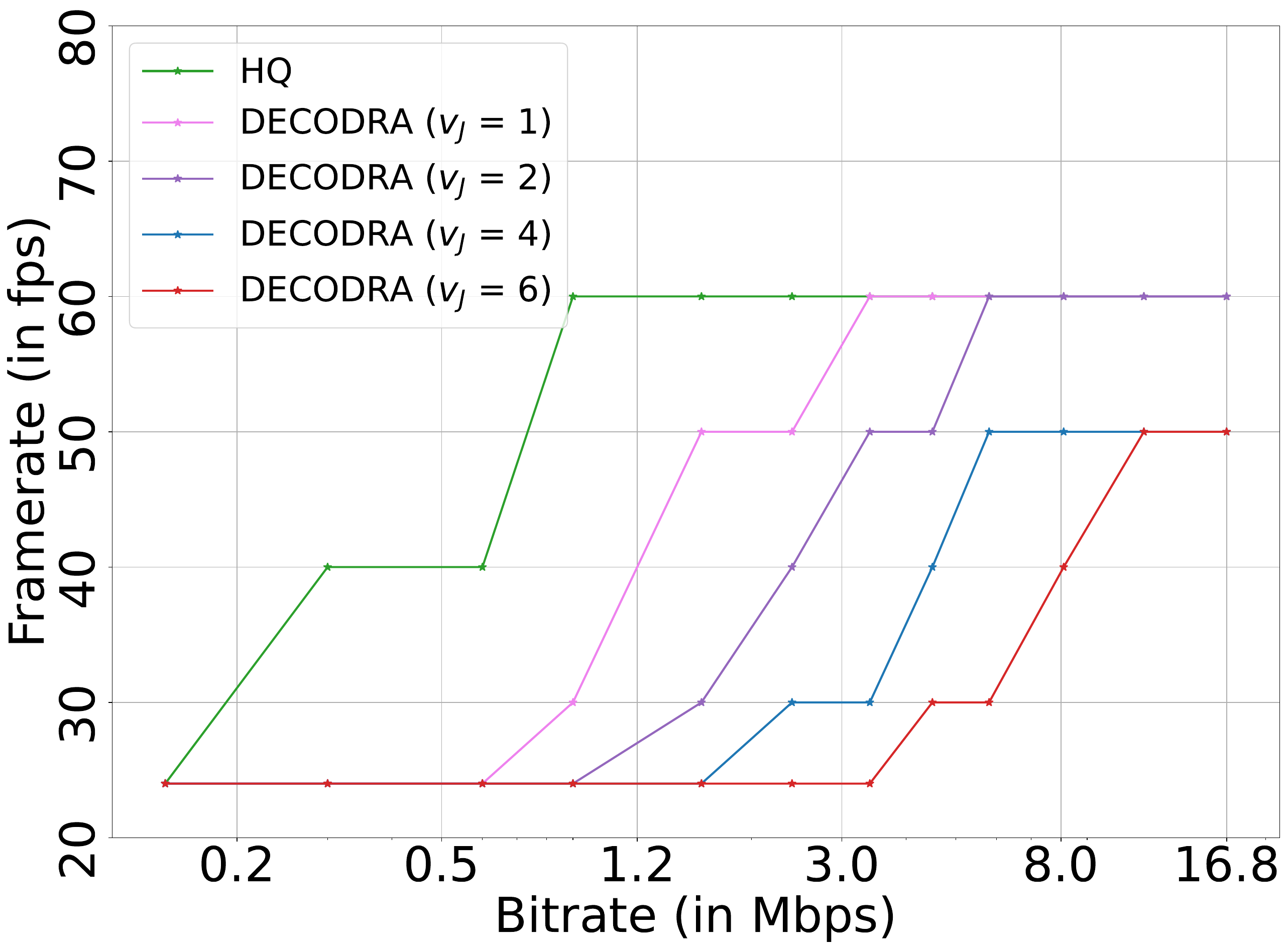}
    \includegraphics[clip,width=\textwidth]{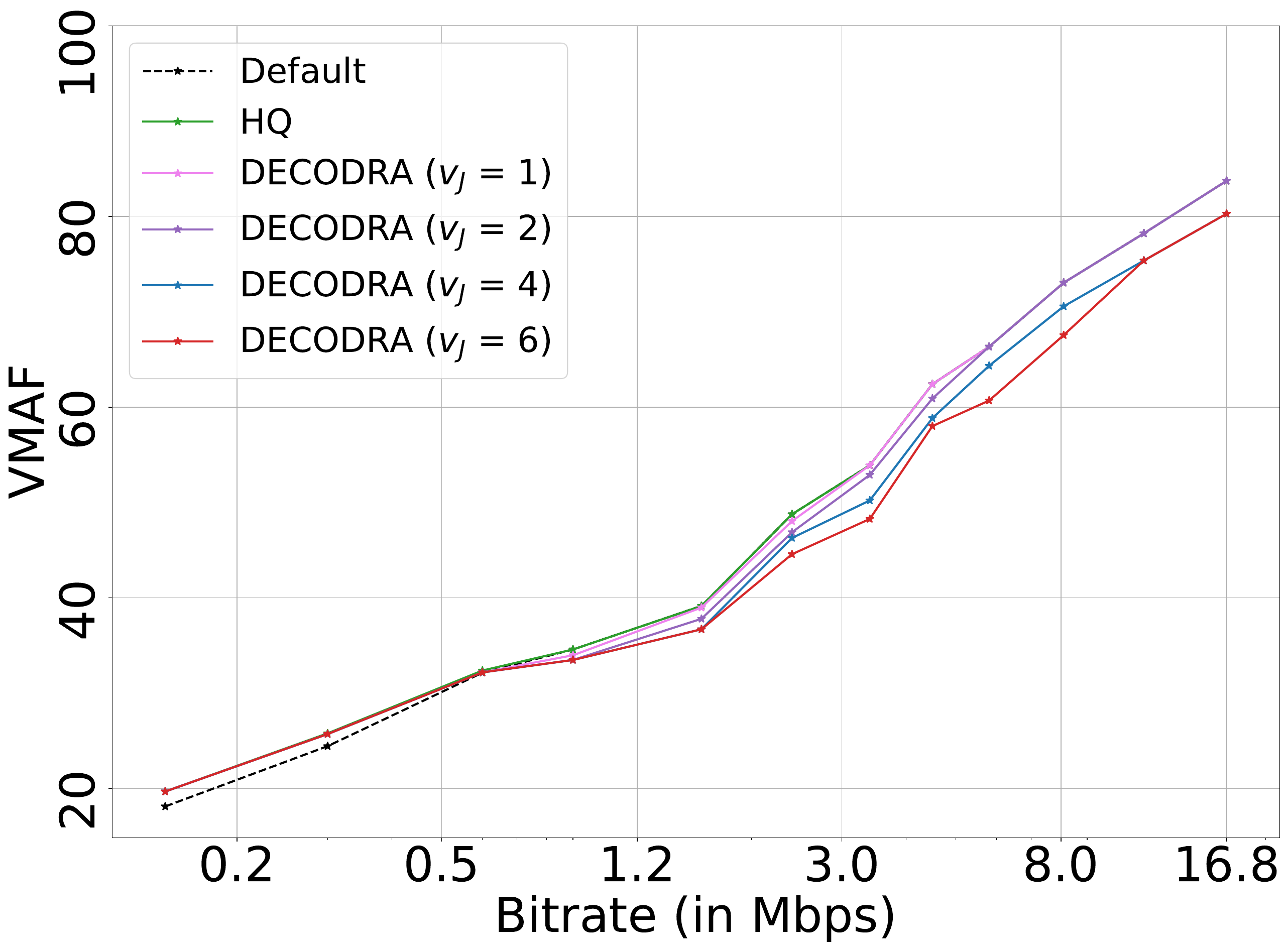}
    \includegraphics[clip,width=\textwidth]{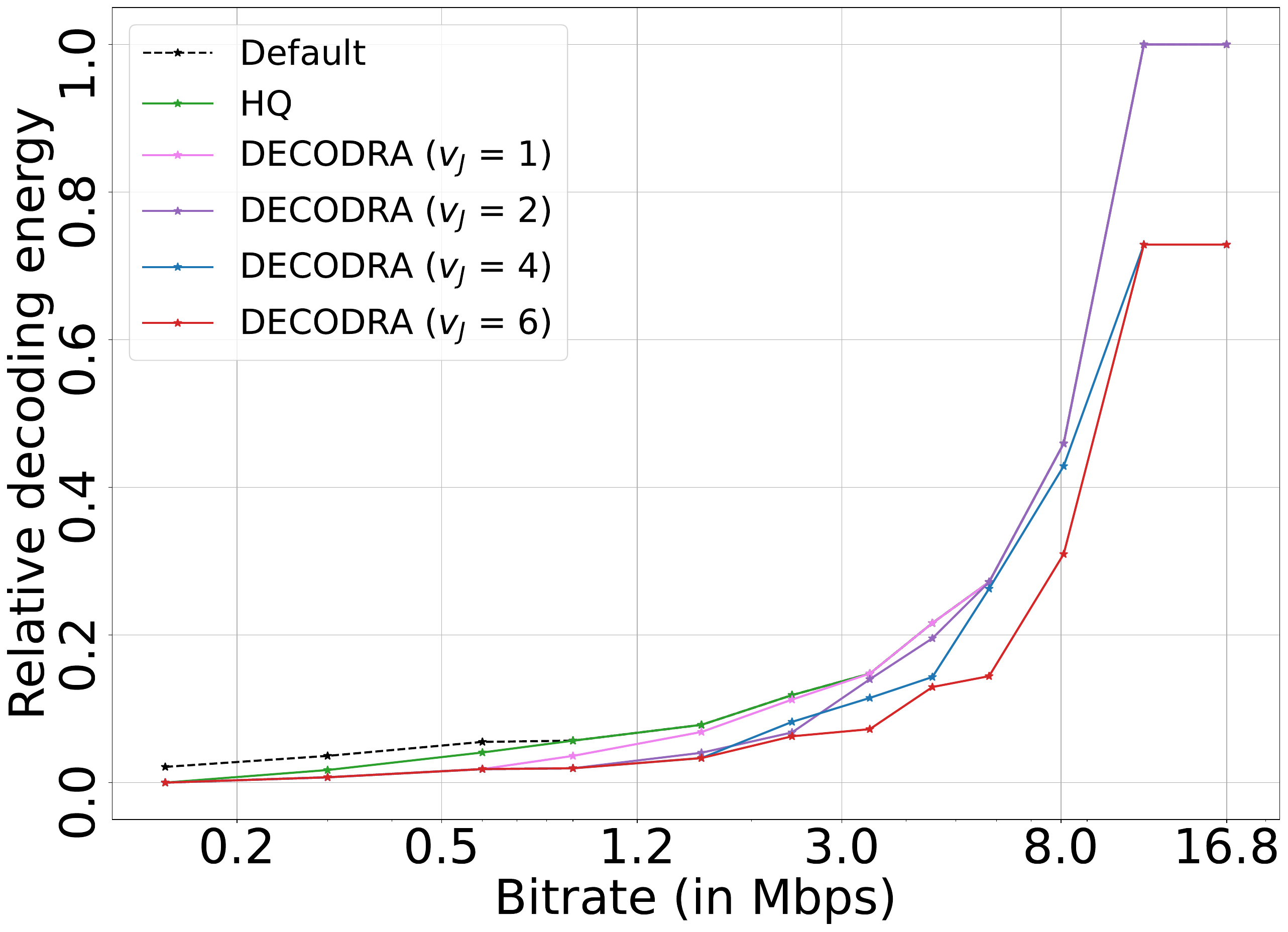}
    \caption{Sequence \textit{0250} (\EY=\num{65.89}, \h=\num{16.57}, \LY=\num{105.39})}
\end{subfigure}
\hfill
\begin{subfigure}{0.491\columnwidth}
    \centering
    \includegraphics[clip,width=\textwidth]{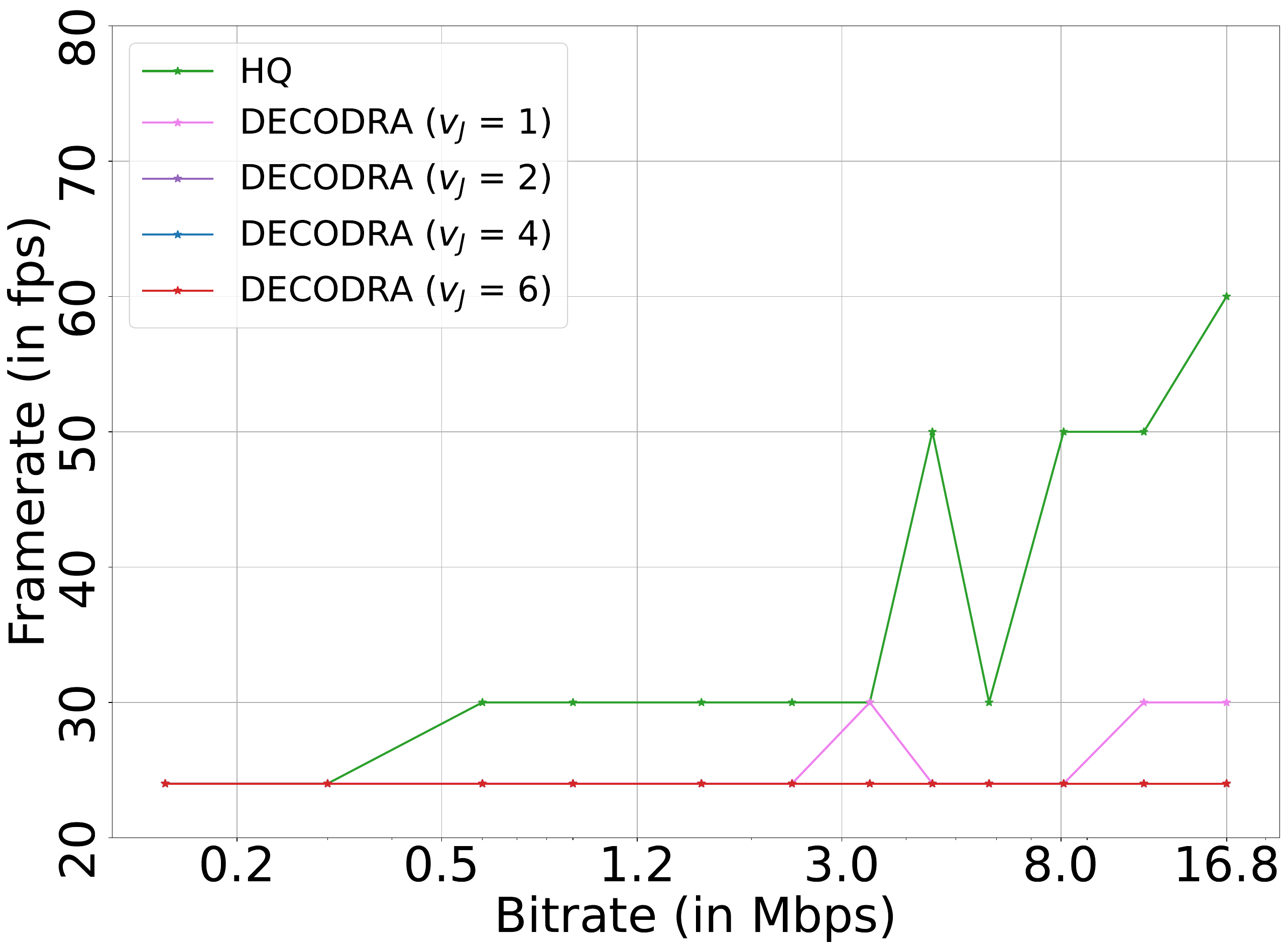}
    \includegraphics[clip,width=\textwidth]{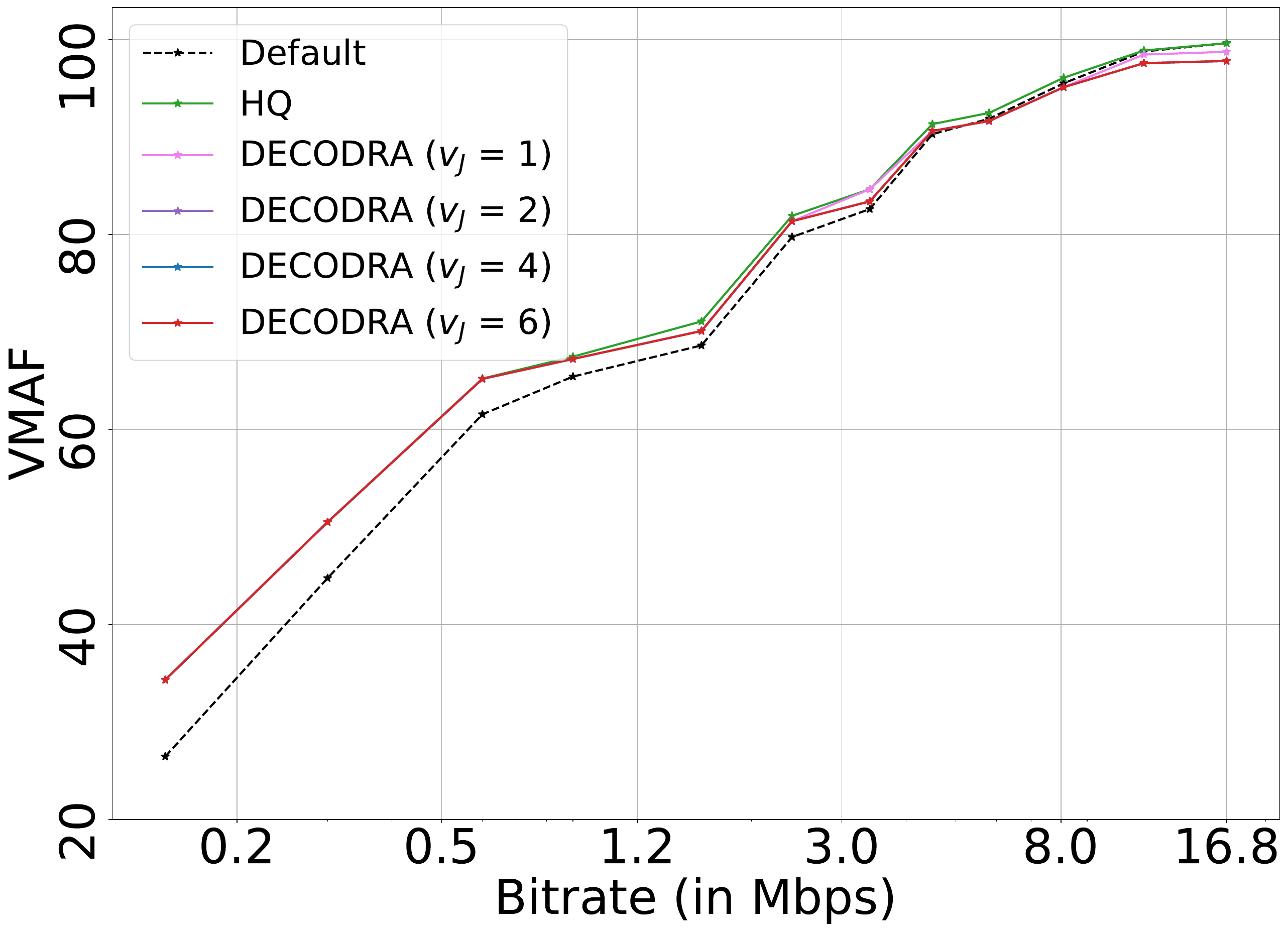}
    \includegraphics[clip,width=\textwidth]{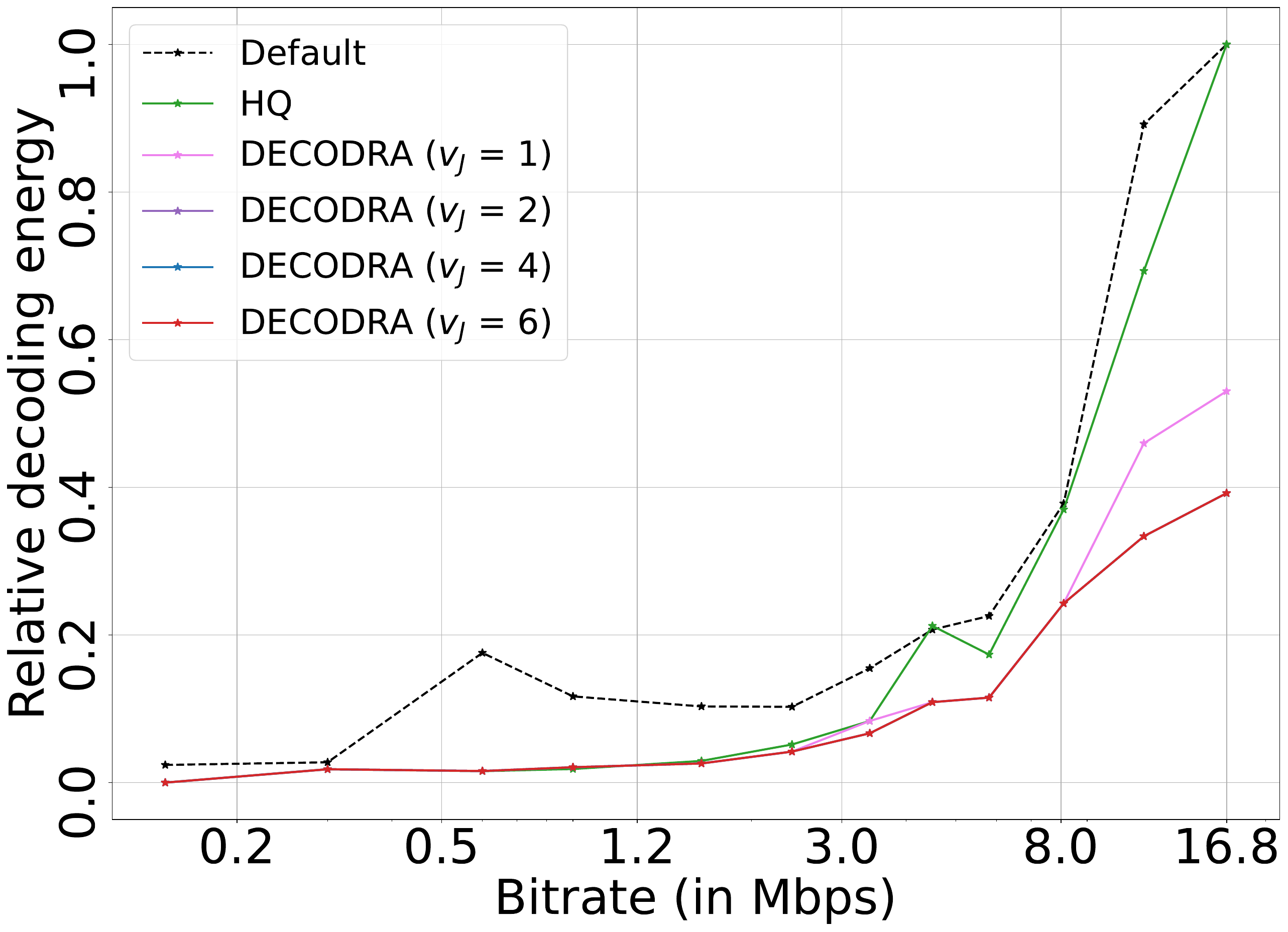}
    \caption{Sequence \textit{0262} ($E_{\text{Y}}$=\num{19.1}, $h$=\num{1.12}, $L_{\text{Y}}$=\num{59.34})}
\end{subfigure}
%\vspace{-0.3em}
\caption{Framerate decision, rate-VMAF, and rate-decoding energy curves of representative sequences of Inter-4K dataset for the \emph{Default}, \emph{HQ}, and \scheme.}
\label{fig:rd_res}
\end{figure}

\section{Experimental results}
\label{sec:results}
We observe in Fig.~\ref{fig:rd_res} that \emph{HQ} generally operates at higher framerates across a range of bitrates, particularly above 3\,Mbps, maintaining close to 60\,fps. \scheme adjust the framerate dynamically, especially at lower bitrates. Higher values of $v_{\text{J}}$, (\eg $v_{\text{J}}$=6) tend to reduce the framerate more aggressively, specifically at lower bitrates. \emph{Default} and \emph{HQ} deliver high VMAF scores, with \emph{HQ} slightly outperforming \emph{Default} in terms of perceptual quality. \scheme achieves similar VMAF to \emph{Default} and \emph{HQ} at higher bitrates (above 3\,Mbps), but as $v_{\text{J}}$ increases (particularly $v_{\text{J}}$=6), there is more quality degradation at lower bitrates (below 1\,Mbps). \scheme provides substantial decoding energy savings at lower bitrates, with more aggressive energy reductions observed for higher $v_{\text{J}}$ values (\eg $v_{\text{J}}$=6 shows the largest energy savings). At higher bitrates (above 3\,Mbps), all methods converge in terms of decoding energy consumption, although \scheme still maintains some energy efficiency advantages. \emph{HQ} and \emph{Default} exhibit similar and relatively higher decoding energy consumption compared to \scheme, especially at lower bitrates.

\begin{figure}[t]
\centering
    \includegraphics[width=0.491\columnwidth]{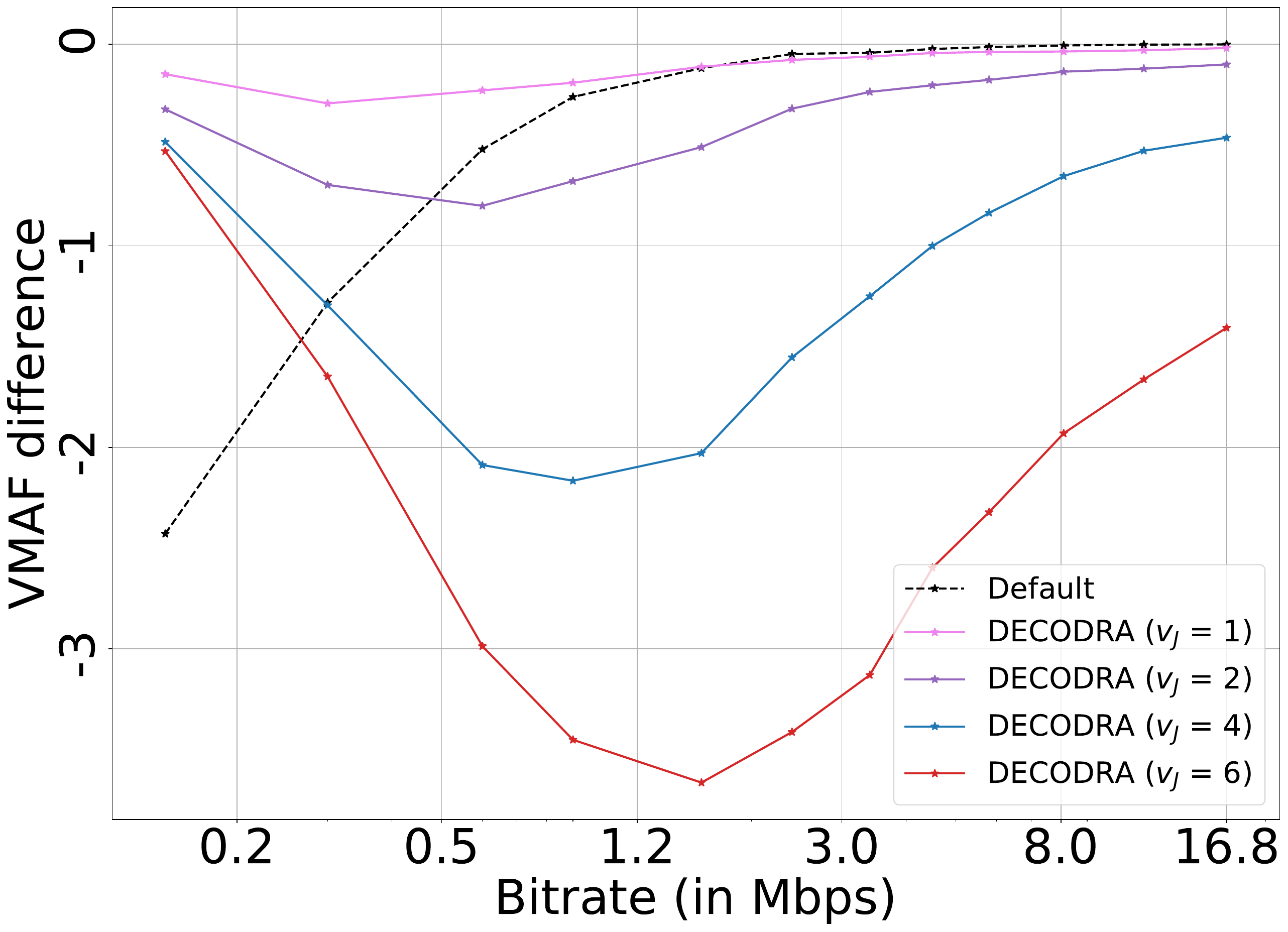}
    \includegraphics[width=0.491\columnwidth]{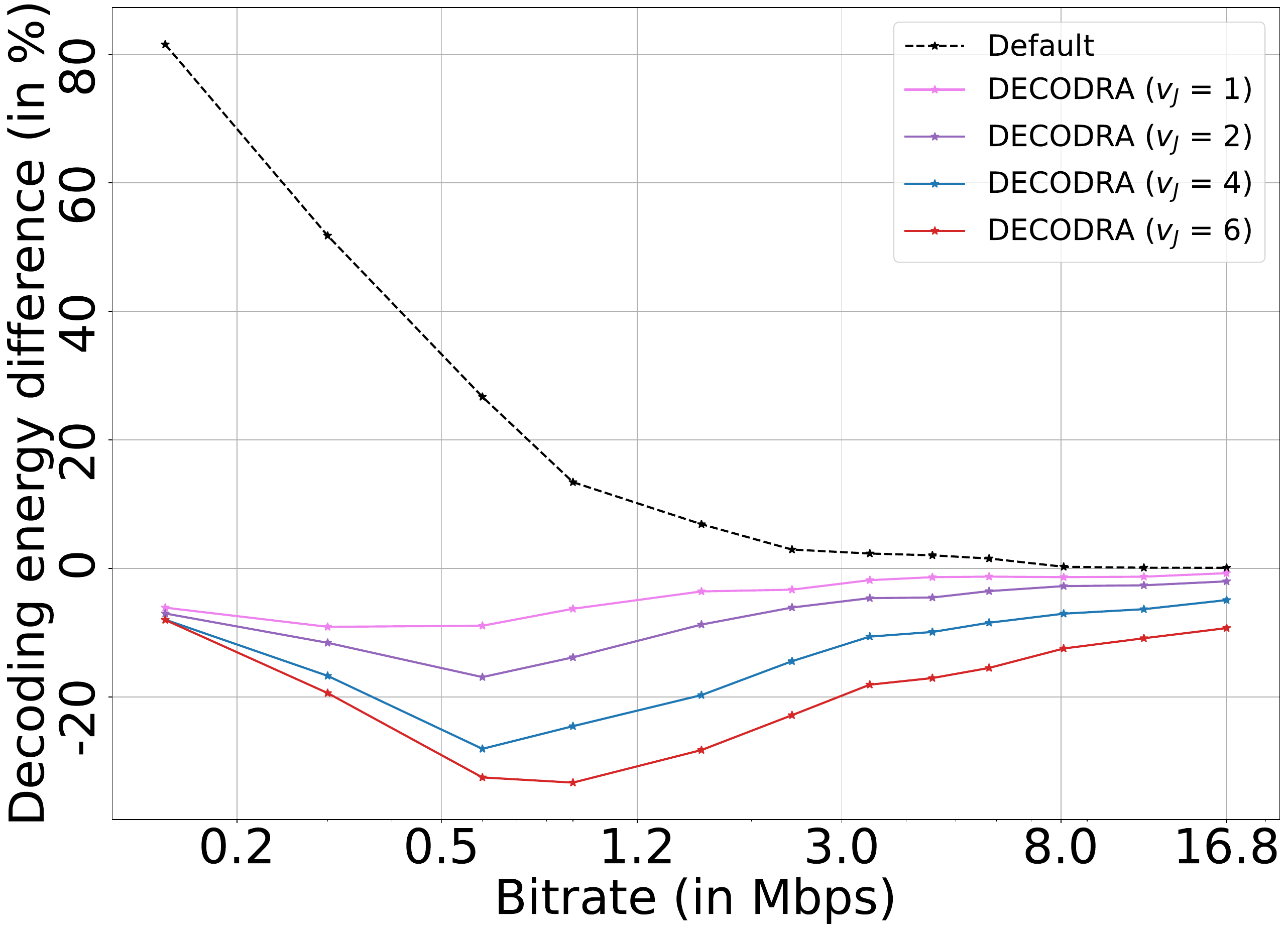}    
\caption{Average VMAF decrease and decoding energy reduction observed for each representation compared to \textit{HQ} for the Inter-4K dataset. %A method is deemed better if it yields a lower VMAF decrease and a higher decoding energy decrease.
}
%\vspace{-0.1em}
\label{fig:heatmap}
\end{figure}

Fig.~\ref{fig:heatmap} shows that \emph{Default} retains more video quality (less VMAF degradation) at low bitrates but consumes more energy compared to \emph{HQ}, especially in the lower bitrate range. It performs similarly to \emph{HQ} at higher bitrates for both quality and energy consumption. \scheme yields greater energy savings but at the cost of higher VMAF degradation (especially at lower bitrates) with more aggressive settings (higher $v_{\text{J}}$). \scheme ($v_{\text{J}}$=1) provides a good balance between energy savings and maintaining video quality, showing modest quality degradation and decent energy savings compared to \emph{HQ}. Thus, this figure highlights that \scheme can significantly reduce decoding energy consumption while offering a tunable balance between video quality and energy savings, depending on the $v_{\text{J}}$ setting. 

Table \ref{tab:res_cons_energy} compares average coding performance across methods relative to \emph{Default} encoding. All methods incur an additional encoding energy consumption of \SI{304.55}{\percent}. Predicting the optimized framerate, rather than relying on brute force encoding as in~\cite{dec_comp_vvc_ref, qadra_ref}, can significantly reduce this computational overhead on the encoding server, which will be explored in future work. \emph{HQ} achieves the best perceptual quality (\mbox{BD-VMAF} of 0.47) but provides only a small reduction in decoding energy. \scheme provides a flexible trade-off between quality and energy, depending on the value of $v_{\text{J}}$. As $v_{\text{J}}$ increases (from 1 to 6), there is a gradual degradation in both PSNR and VMAF, but the method achieves progressively better decoding energy savings, making it more suitable for low-power playback scenarios. $v_{\text{J}}$=6 delivers the best decoding energy reduction (\SI{-13.45}{\percent}) but sacrifices more quality, whereas $v_{\text{J}}$=1 strikes a good balance with minor quality degradation and a reasonable reduction in decoding energy (\SI{-3.22}{\percent}). \scheme ($v_{\text{J}}$=6) maintains the best energy-efficiency trade-off for VMAF with \mbox{BDDE} of \SI{-20.69}{\percent}.
 
\begin{table}[t]
\caption{Average coding performance compared to the \textit{Default} encoding.}
\centering
\resizebox{0.969\columnwidth}{!}{
\begin{tabular}{@{}l@{ }||@{ }c@{ }|@{ }c@{ }|@{ }c@{ }|@{ }c@{ }|@{ }c@{ }|@{ }c@{ }|@{ }c@{ }}
\specialrule{.12em}{.05em}{.05em}
\specialrule{.12em}{.05em}{.05em}
Method & $v_{\text{J}}$  & BD-PSNR & BD-VMAF & $\Delta E_{\text{enc}}$ & $\Delta E_{\text{dec}}$ & BDDE (PSNR) & BDDE (VMAF) \\
& & [dB] & & [\%] & [\%] & [\%] & [\%] \\
\specialrule{.12em}{.05em}{.05em}
\specialrule{.12em}{.05em}{.05em}
%\textit{Default (24fps)} & - & -4.95 &	-11.84 & -37.94 & 	-53.76 &	87.02 &	-26.37 \\
%\textit{Default (30fps)} & - & -3.41 &	-8.19 &	-29.30 & -45.66 &		35.15 &	-22.87 \\
%\textit{Default (48fps)} & - & -1.20 &	-2.81 &	-10.84 &	-8.46 &	21.46 &	-15.09 \\
%\hline
\emph{HQ} & - & \textbf{-0.17} &	\textbf{0.47} & 304.55 &	-1.58 &		-4.71 &	-10.40 \\
\hline
\multirow{4}{*}{\scheme} & 1 & -0.39 &	0.33 &	\multirow{4}{*}{304.55} & -3.22 & 	\textbf{-5.41} & -14.56 \\
 &  2  &  -0.61 &	0.01 &	 & -4.94 & -5.11 &	-16.95  \\
 &  4  & -1.09 &	-0.93 &	 & -8.99  &	-2.33 &	-19.78 \\
 &  6  & -1.59 &	-2.11 &	 & \textbf{-13.45} & 	2.91 &	\textbf{-20.69} \\
\specialrule{.12em}{.05em}{.05em}
\specialrule{.12em}{.05em}{.05em}
\end{tabular}}
\label{tab:res_cons_energy}
\end{table}

\section{Conclusions}
\label{sec:concl}
This paper introduced \scheme, a methodology for optimizing decoding energy consumption at the client devices by dynamically adjusting framerate to meet specified quality degradation thresholds. Our VFR PF estimation minimizes framerate while maintaining acceptable quality. Results demonstrate \scheme's effectiveness across diverse video content, balancing bitrate, visual quality, and decoding energy efficiency. \scheme yields decoding energy reductions of up to \SI{13.45}{\percent} while limiting the VMAF degradation to as little as \SI{0.33}{} points. Even at the most aggressive settings, \scheme maintains acceptable quality with only a 2.11-point VMAF reduction, making \scheme valuable for streaming and mobile video playback applications.

%Future research could explore advanced machine learning models for real-time prediction of quality degradation and decoding complexity. Another potential direction is integrating emerging technologies like machine learning-based content adaptation and network-aware streaming algorithms. Additionally, conducting subjective evaluations across different platforms, content types, and user demographics would further validate our approach.

\balance
\newpage
\bibliographystyle{IEEEtran}
\bibliography{references.bib}
%{\linespread{0.7}\selectfont\bibliography{references.bib}}
\balance
\end{document}